\documentclass[11pt,oneside,letterpaper]{article}
\usepackage{amssymb}
\usepackage{amsmath}
\usepackage[dvips]{graphicx}
\usepackage{setspace}
\usepackage{fancyhdr}
\usepackage{ifpdf}
\usepackage{graphicx}
\usepackage{comment}

\newcommand{\bea}{\begin{eqnarray}}
\newcommand{\eea}{\end{eqnarray}}
\newcommand{\be}{\begin{equation}}
\newcommand{\ee}{\end{equation}}

\numberwithin{equation}{section}
\allowdisplaybreaks  

\begin{document}

\vspace*{2.5cm}
\begin{center}
{ \LARGE \textbf{Hopfing and Puffing Warped Anti-de Sitter Space}\\}
\vspace*{1.7cm}
\begin{center}
Dionysios Anninos
\end{center}
\end{center}
\vspace*{0.6cm}
\begin{center}
Center for the Fundamental Laws of Nature\\
Jefferson Physical Laboratory, Harvard University, Cambridge, MA, USA\\
\vspace*{0.8cm}
\end{center}
\vspace*{1.5cm}
\begin{abstract}
\noindent
Three dimensional spacelike warped anti-de Sitter space is studied in the context of Einstein theories of gravity and string theory, where there is no gravitational Chern-Simons term in the action. We propose that it is holographically dual to a two-dimensional conformal field theory with equal left and right moving central charges. Various checks of the central charges are offered, based on the Bekenstein-Hawking entropy of the stretched warped black holes and warped self-dual solutions. The proposed central charges are applied to compute the Bekenstein-Hawking entropy of the Hopf T-dual of six-dimensional dyonic black strings which have a near horizon consisting of three dimensional warped anti-de Sitter space times a three-sphere. We find that the Hopf T-duality is a map between thermal states with equal entropy of the CFTs dual to the dyonic black string and the Hopf T-dualized black string.

\end{abstract}

\newpage
\setcounter{page}{1}
\pagenumbering{arabic}

\section{Introduction and Summary}

Three dimensional spacelike warped anti-de Sitter space ($WAdS_3$) is the geometry obtained by  expressing $AdS_3$ as a Hopf fibration over $AdS_2$, where the fiber is the real line \cite{Bengtsson:2005zj,Duff:1998cr,Strominger:1998yg}, and ``warping" the length of the fiber. Its metric is given by:
\be
ds^2 = \frac{(4-\beta^2)\ell^2}{12}\left[ -\cosh^2\rho d\tau^2 + d\rho^2 + \beta^2 (dz^2 + \sinh\rho d\tau)^2 \right] \label{intro}
\ee
where $z \in [-\infty,\infty]$ is the fiber coordinate and $\beta^2 \in [0,4]$ is the warp factor.

When the fiber is stretched ($\beta^2 > 1$) we have \emph{stretched} $WAdS_3$, whereas when the fiber is squashed ($\beta^2<1$) we have \emph{squashed} $WAdS_3$. In \cite{Anninos:2008fx}, these solutions and their corresponding black holes were studied in the context of topologically massive gravity (TMG) \cite{Deser:1981wh,Deser:1982vy}. It was proposed that quantum gravity in spacelike stretched $WAdS_3$ is holographically dual to a two-dimensional CFT and a left and right moving central charge was proposed.

In this note, we wish to study the warped black holes and warped self-dual objects discussed in \cite{Anninos:2008fx} in the context of Einstein theories of gravity and string theory. In these theories there is no gravitational Chern-Simons (CS) term in the three dimensional effective Lagrangian. We examine the known cases where such solutions arise in theories of Einstein gravity with matter content. We propose that in such contexts both squashed and stretched spacelike $WAdS_3$ is given by a two-dimensional CFT with equal left and right-moving central charges given by
\bea
c_L = c_R = \frac{\sqrt{3}(4-\beta^2)^{1/2} \beta \ell}{2G_3}
\eea
where $-6/\ell^2$ is the Ricci scalar and $\beta^2$ is the warp factor of the warped solution \ref{intro}.

The central charges are obtained by studying the Bekenstein-Hawking entropy of stretched black holes and of both stretched and squashed warped self-dual solutions. These central charges only depend on the warp factor and the curvature scale, which are both geometrical quantities. Furthermore, when the solution becomes unwarped with $\beta^2 = 1$ the central charges reduce to the ones for $AdS_3$. Finally, one can express the Bekenstein-Hawking entropy of the warped black holes and self-dual solutions in the form of the entropy of a thermal state in a two-dimensional CFT by using the above central charges. The left and right-moving temperatures of the thermal state are related to the coefficients of the discrete identifications made on $WAdS_3$ to obtain the warped black holes. One way to confirm these central charges would be to find the appropriate boundary conditions for warped anti-de Sitter space and study the asymptotic symmetry group (ASG), in the same spirit as Brown-Henneaux \cite{Brown:1986nw}.

The proposed central charges are then applied to match the Bekenstein-Hawking entropy of the black objects with $WAdS_3\times S^3$ near horizons, discovered in \cite{Duff:1998cr}, with the entropy of the proposed dual CFT\footnote{It worth noting that both stretched and squashed $WAdS_3$, with the fiber coordinate identified, can also appear as the near horizon geometry of the extremal Kerr black hole at fixed polar angle \cite{Bengtsson:2005zj,Bardeen:1999px,ghss,sstrings}.}. These objects are Hopf T-duals of the dyonic string \cite{Duff:1995yh} in type IIB string theory reduced on a $T^4$ or $K3$. A Hopf T-dual solution is obtained by compactifying the fiber coordinate $z$ of the six-dimensional type IIB solution, performing a T-duality along $z$ and finally oxidizing back to six-dimensions. The type IIB dyonic string sources both NS-NS and R-R charges and has an $AdS_3\times S^3$ near horizon.
Using the fact that $AdS_3$ is dual to a two-dimensional CFT, we extend the duality chain obtained in \cite{Duff:1998cr} to the following:
\be
CFT_B \longleftrightarrow AdS_3 \times S^3 \longleftrightarrow WAdS_3 \times S^3 \longleftrightarrow CFT_A
\ee
where both $AdS_3$ and $WAdS_3$ have the fiber coordinate identified. We compute the entropy of the thermal states of the CFTs on either side using the proposed central charges and find that they match, in agreement with the observation that the entropy of the black string solutions remains unchanged under a Hopf T-duality \cite{Duff:1998cr}. Thus, the duality maps thermal states with vanishing right-moving temperature in $CFT_B$ to thermal states with vanishing right-moving temperature in $CFT_A$ with the same entropy.

Finally, the duality between $AdS_3\times S^3$ and the theory obtained by Hopf T-dualizing along the fiber coordinate of the $S^3$ is studied. We find that the Hopf T-duality is a map between thermal states with vanishing right-moving temperature with the same entropy. The CFTs on either side have the same central charge. A similar conclusion is obtained for the duality between $WAdS_3\times S^3$ and theory obtained by Hopf T-dualizing along the fiber coordinate of the $S^3$.

\section{Black Hole Entropies}

Our story begins by studying the entropies of both the BTZ and warped black holes in the context of Einstein theories of gravity. As usual the entropy is given by the Bekenstein-Hawking formula
\be
S_{BH} = \frac{A_{BH}}{4G_3}
\ee
where $A_{BH}$ is the proper area of the black hole and $G_3$ is the three-dimensional Newton's constant.

\subsection{$AdS_3$ and BTZ Black Holes}

We begin with the metric of global $AdS_3$ expressed as a Hopf fibration of the real line over Lorentzian $AdS_2$:
\be
ds^2 = \frac{\ell^2}{4}\left[ -\cosh^2\rho d\tau^2 + d\rho^2 + (dz + \sinh\rho d\tau)^2 \right]
\ee
where $z$ is known as the fiber coordinate and $\{\rho, \tau, z\} \in [-\infty,\infty]$. The isometry group is given by $SL(2,\mathbb{R})_R\times SL(2,\mathbb{R})_L$ and the explicit form of the Killing vectors is given in appendix \ref{appa}.

As is well known by now, BTZ black holes arise in Einstein gravity endowed with a negative cosmological constant $\Lambda = -1/\ell^2$ \cite{Banados:1992wn}. Furthermore, these black holes have been shown to be global identifications of $AdS_3$ \cite{Banados:1992gq,Maldacena:1998bw,Carlip:1994gc}\footnote{In what follows we will be focusing on non-extremal warped and unwarped black holes.}. Specifically, they are obtained by periodically identifying points along Killing directions of $AdS_3$ and excluding the region of space where CTCs arise (behind the horizon). The identified Killing direction is given by:
\be
\xi = \pi\ell\left( \tilde{T}_L J_2 - \tilde{T}_R \tilde{J}_2 \right)
\label{id}
\ee
where $\tilde{T}_L = (r_+ + r_-)/2\pi\ell^2$ and $\tilde{T}_R = (r_+ - r_-)/2\pi\ell^2$ are the corresponding left and right moving temperatures of the dual theory and $r_+$ and $r_-$ are the outer and inner horizon radii.

The central charges of the dual CFT were discovered by Brown-Henneaux \cite{Brown:1986nw}:
\be
c_L = c_R = \frac{3\ell}{2G_{3}}
\label{cbh}
\ee
Finally, the Bekenstein-Hawking entropy of the BTZ black hole is given by
\be
S_{BH}^{BTZ} = \frac{\pi r_+ }{2G_3} = \frac{\pi^2 \ell}{3}\left( \tilde{T}_L c_L + \tilde{T}_R c_R \right)
\ee
The entropy takes the form of a Cardy-formula in accordance with the notion that the theory is dual to a two-dimensional CFT.

\subsection{$WAdS_3$ and Warped Black Holes}

In \cite{Anninos:2008fx}, $WAdS_3$ was obtained and studied as a solution to pure TMG \cite{Deser:1981wh,Deser:1982vy}. The explicit metric for spacelike $WAdS_3$ has the form:
\be
ds^2 = \frac{(4-\beta^2)\ell^2}{12}\left[ -\cosh^2\rho d\tau^2 + d\rho^2 + \beta^2 (dz^2 + \sinh\rho d\tau)^2 \right]
\ee
where the Ricci scalar is given by $R = -6/\ell^2$ and $\{\rho,\tau,z\} \in [-\infty,\infty]$. The relation to the TMG parameter $\nu$ is given by
\be
\beta^2 = \frac{4\nu^2}{(\nu^2 + 3)}
\ee
Notice that the whole range of $\nu^2 \in [0,\infty]$ spans the interval $\beta^2 \in [0,4]$. We will restrict $\beta^2$ to this range in what follows.

The isometry group of $WAdS_3$ is given by $SL(2,\mathbb{R})_R\times U(1)$, where the $U(1)$ is spacelike. The warp factor is given by $\beta^2$ such that when $\beta^2>1$ we have a \emph{stretched} warped solution, and when $\beta^2<1$ we have a \emph{squashed} warped solution.

The metric of the spacelike warped black hole discovered in \cite{Moussa:2003fc,Bouchareb:2007yx}, is based on the solutions obtained in \cite{Nutku:1993eb,Gurses} and given by
\begin{multline}
\frac{ds^2}{\ell^2}= {dt^2}+\frac{ (4-\beta^2) d{r}^2}{12 ({r}-{r}_{+})({r}-{r}_{-})} + \frac{2\sqrt{3}}{\sqrt{4-\beta^2}}\left(\beta {r}  - \sqrt{{r}_{+}{r}_{-}}\right)dtd\theta \\
+\frac{3{r}}{(4-\beta^2)}\left((\beta^2 - 1){r}+({r}_+ + {r}_-) - 2\beta\sqrt{{r}_+{r}_-}\right)d\theta^2
\label{eq:ng}
\end{multline}
Notice that squashed black holes, with $\beta^2 < 1$, have CTCs close to the boundary where $r$ is large. Thus, it is unclear whether squashed black holes are part of the physical spectrum of the theory.

It was further shown in \cite{Anninos:2008fx} (see also \cite{Moussa:2008sj}) that warped black holes are global identifications of $WAdS_3$ under a linear combination of the Killing vectors of $WAdS_3$:
\be
\xi = \pi\ell\left( T_L J_2 - T_R \tilde{J}_2 \right) \label{id2}
\ee
where
\bea
T_{L} &\equiv& {3\over2\pi\ell(4-\beta^2)}\left({r}_+ + {r}_- -
{2\over\beta}\sqrt{{r}_+{r}_-}\right)\label{tl}\\
T_{R} &\equiv& {3({r}_+-{r}_-)\over2\pi\ell(4-\beta^2)} \label{tr}
\eea
Here one interprets the coefficients as the left and right moving temperatures, as is done for the BTZ black hole.

\subsubsection{Warped Black Holes in TMG}

The proposed central charges for the dual CFT in the context of TMG were conjectured in \cite{Anninos:2008fx} to be\footnote{Recently, $c_R$ in \ref{ctmg} has been obtained by studying the ASG in \cite{Compere:2008cv}.}:
\bea
\tilde{c}_R &=& \frac{(5\nu^2 + 3)\ell}{G_3 \nu (\nu^2+3)} = \frac{(1+\beta^2)(4-\beta^2)^{1/2}\ell}{G_3\sqrt{3}\beta} \\
\tilde{c}_L &=& \frac{4\nu \ell}{G_3(\nu^2+3)} = \frac{\beta(4-\beta^2)^{1/2}\ell }{G_3\sqrt{3}}.\label{ctmg}
\eea
Notice that they are not equal, implying a diffeomorphism anomaly in the boundary CFT. In fact, the diffeomorphism anomaly arises from the gravitational CS term and was shown to agree with that predicted from the bulk point of view in \cite{Kraus:2005zm}:
\be
\tilde{c}_L - \tilde{c}_R = -\frac{\ell}{G_3\nu}
\ee
The CS corrected entropy \cite{Solodukhin:2005ah,Tachikawa:2006sz,Kraus:2005vz,Park:2006gt} for the warped black hole in TMG can be written as \cite{Anninos:2008fx}
\be
S^{WBH}_{TMG} = \frac{\pi^2 \ell}{3}\left( T_L \tilde{c}_L + T_R \tilde{c}_R \right)
\ee
which once again takes the form of the entropy for a two-dimensional CFT.
\subsubsection{Warped Black Holes in Einstein Gravity}

We will be interested in the central charges relevant to the stretched warped black hole in Einstein theories of gravity. We begin by reviewing the known cases where $WAdS_3$ appears in Einstein theories of gravity with matter content.

\subsubsection*{Perfect Fluids}
It was shown in \cite{Gurses} that solutions to TMG with a cosmological constant $\Lambda$ can be obtained from solutions to Einstein gravity coupled to a stress tensor that takes the form of a perfect fluid. That is to say
\be
R_{\mu \nu} -\frac{1}{2}g_{\mu \nu} R = T^{matter}_{\mu \nu}
\ee
where
\be
T^{matter}_{\mu\nu}  = (p + \rho)u_\mu u_\nu + \rho g_{\mu \nu}
\ee
The pressure and energy density of the fluid are denoted by $p$ and $\rho$ such that the cosmological constant in the TMG theory is $\Lambda = (2p - \rho)/3 = -1/\ell^2$. In fact for such matter content to support stretched black holes in \ref{eq:ng} one requires $u_\mu u^\mu = +1$, i.e. a spacelike four-velocity, for the perfect fluid. The warp factor is given by $\beta^2 = 4p\ell^2/(p\ell^2+3)$.

\subsubsection*{Topologically Massive Electrodynamics}
Squashed warped black holes, which have CTCs, arise in topologically massive electrodynamics coupled to Einstein gravity and a negative cosmological constant \cite{Andrade:2005ur,Banados:2005da} with action
\be
I_{TME} = \frac{1}{16\pi G}\int d^3 x \left[ \sqrt{-g} \left( R + \frac{2}{\ell^2} - \frac{1}{4}F^2 \right) - \frac{\alpha}{2} A_\mu F_{\nu \rho}\varepsilon^{\mu\nu\rho}\right]
\ee
Stretched warped black holes can also arise in such theories, with a warp factor $\beta^2 = 2\alpha^2\ell^2/(\alpha^2\ell^2+1)$, but they require wrong sign kinetic terms for the Maxwell field strength. It was shown in \cite{Banados:2005da} that the standard black hole thermodynamics holds for warped black holes in topologically massive electrodynamics using the Bekenstein-Hawking form of the entropy.

\subsubsection*{Near horizon of a Extremal Kerr black hole}
$WAdS_3$ is revealed to be the near horizon of extremal Kerr black holes at fixed polar angle \cite{Bengtsson:2005zj,Bardeen:1999px}\footnote{In fact the fiber coordinate is periodically identified, so it is actually a self-dual solution as discussed in section \ref{sd}.}. Interestingly, it also arises as the external metric to an extremely relativistic rotating disk at fixed polar angle \cite{bardeen} given a specific limit. The metric is given by \cite{Bardeen:1999px},
\be
ds^2 = 2 M^2 \Omega^2 \left( -(1+r^2)d\tau^2 + \frac{dr^2}{(1+r^2)} + d\theta^2 + \beta(\theta)^2(d\phi^2 + r d\tau)^2 \right)
\label{kerr}
\ee
where the warp factor and $\Omega$ are given by,
\be
\beta(\theta) = \frac{2\sin\theta}{1+\cos^2\theta}, \quad \Omega^2 = \frac{1 + \cos^2\theta}{2}\label{kerrpars}
\ee
The coordinate $\phi$ has a $2\pi$ periodicity, $\theta\in[0,\pi]$ and $r\in[-\infty,\infty]$. The angular momentum is given in terms of the mass $M$ and $G_4$ by $J = M^2/G_4$.

\subsubsection*{Uplifted Solutions in String Theory}
Squashed $WAdS_3$ can be lifted to a full string theory solution in ten dimensions, as shown in \cite{Compere:2008cw}. Also, black strings with near horizons given by $WAdS_3\times S^3$ were shown to be Hopf-T duals of the dyonic black string in six-dimensions \cite{Duff:1998cr}, supported by both NS-NS and R-R charges, which has an $AdS_3 \times S^3$ near horizon. We will return to this case in section \ref{dyo}. Finally squashed $WAdS_3$ has appeared as an exact string background obtained from the target space of an exact marginal deformation of the $SL(2,\mathbb{R})$ WZW world sheet model \cite{Detournay:2005fz,Israel:2003ry,Israel:2004vv}.

\subsection{Entropy of Warped Black Holes and Central Charges}

For Einstein theories of gravity, the metric \ref{eq:ng} has a Bekenstein-Hawking entropy is given by the area of the black hole divided by $4G_3$. Using the left and right moving temperatures given in \ref{tl} and \ref{tr} this gives\footnote{The relation \ref{es} has also been noticed in the context of squashed black holes in \cite{Compere:2008cw}.}
\bea
S_{BH}^{WBH} &=& \frac{\sqrt{3}\pi \ell}{2(4-\beta^2)^{1/2} G_3}\left( {\beta} r_+ - \sqrt{r_+ r_-}  \right)\\
    &=& \frac{\pi^2 \ell}{3}\left( \frac{\sqrt{3}(4-\beta^2)^{1/2} \beta \ell}{2G_3} T_L + \frac{\sqrt{3}(4-\beta^2)^{1/2} \beta \ell}{2G_3} T_R  \right)\label{es}
\eea
Thus we propose that in the Einstein theory the central charges of spacelike stretched $WAdS_3$ are given by
\be
c_R = c_L = \frac{\sqrt{3}(4-\beta^2)^{1/2} \beta \ell}{2G_3} 
\label{cw}
\ee
The above central charges pass all the obvious tests. For example, there is no diffeomorphism anomaly for Einstein theories such that the difference between the central charges vanishes. Furthermore, they reduce to those of $AdS_3$, i.e. $3\ell/2G_3$, when $\beta = 1$. Finally, they don't depend on the black hole parameters, $r_+$ and $r_-$. Of course, we should emphasize that these central charges are reliable in the limit where gravity is reliable and the curvature is small compared to $\ell_{Pl}^{-2}$.

One could, in principle, verify these central charges by studying the asymptotic symmetries in the spirit of Brown-Henneaux. This has been done in \cite{Compere:2007in} for squashed black holes, where they obtain the above expression of $c_R$ up to a sign. We should note, however, that it would be ideal to find a context in which we can compute the central charges for stretched black holes, given that it is unclear whether squashed black holes (which have CTCs) form part of the physical spectrum of states in the dual CFT. This is difficult because there is no known action in Einstein gravity, with non-pathological matter content, supporting stretched black holes. If the limit $\beta \to 0$ is allowed, the geometry tends to $AdS_2\times \mathbb{R}$ and the proposed central charges vanish at $\beta = 0$. Finally, if $\beta^2 > 4$ the central charges above become imaginary, however in all cases mentioned, the stretched solutions are at most stretched with $\beta^2 = 4$, where the central charges vanish again.

On the other hand, there is another set of non-pathological solutions obtained by identifying the fiber coordinate which are likely to be part of the physical spectrum - given the near near horizon of the extremal Kerr black hole at fixed polar angle is one of them. In the next section we will study these solutions to get further evidence for $c_L$ for all values of $\beta^2 \in [0,4]$.

\section{Self-Dual Solutions}
\label{sd}

Even though the conjecture in \cite{Anninos:2008fx} was formulated for stretched $AdS_3$, evidence was proposed for the squashed case. Specifically, certain \emph{self-dual} solutions were shown to be physical solutions. These solutions have the fiber coordinate $z$ identified\footnote{There is actually an additional subtlety about squashed solutions in TMG. The limit $\beta \to 0$ is ill defined since the coefficient in front of the CS part of the action blows up. This is not an issue in the Einstein theory, and the limit leads smoothly to an $AdS_2\times \mathbb{R}$ solution where the proposed central charges in \ref{cw} vanish.}. The reason we call such solutions self-dual is that they resemble the self-dual solutions discussed in \cite{Coussaert:1994tu}, where the fiber coordinate is identified for the case of $AdS_3$.

Let us identify the $t$ coordinate in \ref{eq:ng} such that $t \sim t + 2\pi\alpha$. The Bekenstein-Hawking entropy for the warped self-dual solution is given by
\be
S_{BH}^{WSD} = \frac{A_{BH}}{4G_3} = \frac{\pi\alpha\ell}{2G_3}
\ee
It was shown in \cite{Anninos:2008fx} that, based on the identification, one can naturally define a left moving temperature for these solutions as $T_L = \sqrt{3}\alpha/\beta(4-\beta^2)^{1/2}\pi\ell$, while $T_R = 0$. Then we can write our entropy as
\be
S_{BH}^{WSD} = \frac{\pi^2 T_L \ell}{3}\frac{\sqrt{3}(4-\beta^2)^{1/2} \beta \ell}{2G_3} = \frac{\pi^2 \ell T_L c_L}{3} = S^{WSD}_{CFT}\label{sde}
\ee
Thus obtaining another confirmation of $c_L$, which in fact holds even when $\beta^2 < 1$. If the dual theory is a two-dimensional CFT with no diffeomorphism anomaly, knowledge of $c_L$ suffices to conclude that $c_R = c_L$. Thus we have indirect evidence for $c_R$. It would be interesting to learn, for the squashed case, what the non-pathological bulk solutions corresponding to states with both left and right moving temperatures in the CFT are.

The relation \ref{sde} is in fact a property of the self-dual solutions of regular $AdS_3$, which can be confirmed by taking $\beta^2 = 1$. We should note that regular self-dual solutions in $AdS_3$ have the confusing property that the dual boundary theory has closed null curves, once the Killing vectors are conformally rescaled\footnote{One way to get around this is to turn on an right moving temperature $T_R$, such that the solution is a regular BTZ black hole, and take the limit $T_R \to 0$.}. This does not seem to occur for the case of warped self-dual solutions; however, the notion of a conformal boundary becomes less clear in this case \cite{Bengtsson:2005zj}, and it deserves a better understanding. The fact that the Cardy-formula still holds suggests that we take the notion of a dual CFT seriously for these solutions as well. We will now proceed to explore a striking relation between the warped and unwarped self dual solutions discovered in \cite{Duff:1998cr}.

\section{Hopf T-Duality}
\label{dyo}

It was shown in \cite{Duff:1998cr} that there is a type of T-duality transformation that one can perform on solutions of type IIB string theories compactified on a $T^4$ or $K3$ to obtain solutions of type IIA string theory compactified on a $T^4$ or $K3$. In particular they showed that $AdS_3\times S^3$ solutions (or solutions with such near-horizons) of a six-dimensional consistent truncation of type IIB theory is T-dual to a $WAdS_3\times S^3$ solution (or solutions with such near horizons) of a six-dimensional truncation of IIA.

They obtained this by expressing $AdS_3$ as a fibration of the real line over $AdS_2$, compactifying along the fiber coordinate and Hopf T-dualizing such that the length of the fiber is no longer equal to the anti-de Sitter length in the T-dual theory. Thus they showed,
\be
AdS_3\times S^3 \longleftrightarrow WAdS_3 \times S^3
\label{arrow}
\ee

To be more precise, the theories that are dual to each other are in fact $AdS_3\times S^3$ and $WAdS_3 \times S^3$ with the fibrated direction compactified, and are thus the self-dual solutions discussed in section \ref{sd}. At the level of the massless Kaluza-Klein modes, either all or none of the supersymmetry is preserved under a Hopf T-duality depending on the orientation associated with the Hopf reduction \cite{Duff:1998cr}.

\subsection{Solutions}

The matter content of the truncated type IIB theory consists of two dilatons, $\phi_1$ and $\phi_2$, two axions, $\chi_1$ and $\chi_2$, an NS-NS three-form $F^{NS}_{(3)}$ and an R-R three-form $F^{RR}_{(3)}$. The matter content of the IIA theory consists of two dilatons, $\phi_1$ and $\phi_2$, a four-form $F_{(4)}$, a three-form $F_{(3)}$ and a two-form $F_{(2)}$. For the sake of simplicity, all the dilatons are set to zero in what follows. In appendix \ref{appb} we give further details on the construction of the solutions.

\subsubsection*{Type IIB Solutions}
The $AdS_3\times S^3$ solution of the six-dimensional type IIB theory, which arises as the near horizon of the dyonic string \cite{Duff:1995yh} carrying both NS-NS and R-R charges,  is given by the following metric:
\be
ds^2 = \frac{1}{(\lambda^2+\mu^2)}[ - \cosh^2\rho d\tau^2 + d\rho^2 + (dz+\sinh\rho d\tau)^2] + ds^2(S_B^3)
\label{iib}
\ee
and
\bea
F_{(3)}^{NS} &=& \lambda \varepsilon(AdS_3) + \lambda \varepsilon(S_B^3) \\
F_{(3)}^{RR} &=& \mu \varepsilon(AdS_3) + \mu \varepsilon(S_B^3)
\eea
The fiber coordinate $z$ is not necessarily compactified in the above solution. We will compactify it in order to obtain the Hopf T-dual solution in the type IIA theory. The Ricci tensors are
\be
R_{\mu\nu} = -(\lambda^2 + \mu^2)g_{\mu\nu}/2, \quad R_{mn} = (\lambda^2 + \mu^2)g_{mn}/2
\ee
where the Greek indices are those of $AdS_3$ and the Latin indices are those of the $S_B^3$. Thus, the anti-de Sitter length is given by
\be
\ell^2 = \frac{4}{\lambda^2 + \mu^2} = 4(Q_{NS}^2 + Q_{RR}^2)^{1/2}\label{lads}
\ee
where $Q_{NS}$ and $Q_{RR}$ are the NS-NS and R-R charges of the two three-forms present in the truncated type IIB action. Explicitly, they are given by:
\bea
Q_{NS} &\equiv& \frac{1}{16\pi^2}\int_{S^3_B} F^{NS}_{(3)} = \frac{\lambda}{(\lambda^2 + \mu^2)^{3/2}}\\
Q_{RR} &\equiv& \frac{1}{16\pi^2}\int_{S^3_B}  F^{RR}_{(3)}  = \frac{\mu}{(\lambda^2 + \mu^2)^{3/2}}
\eea
Note that distances are measured with respect to the string scale, $\ell_s$. For gravity to be reliable we require a small curvature which equates to having $Q_{NS} \gg 1$ and/or $Q_{RR} \gg 1$.

When we do a Hopf T-duality to obtain the type IIA solution, we compactify along the fiber coordinate $z$, such that $z \sim z + 2\pi$, and perform the field transformations spelled out in \cite{Duff:1998cr}. The six-dimensional string coupling constant is given by
\be
g_6 = e^{(\phi_1 + \phi_2)/2}
\ee
and the radius of the compactification circle in the string frame is given by,
\be
R_{IIB} = e^{(\phi_1 + \phi_2)/4+\sqrt{3/8}\varphi} = \frac{1}{\sqrt{\lambda^2 + \mu^2}} = (Q_{NS}^2+Q_{RR}^2)^{1/4}
\ee
where $\varphi$ is the dilaton corresponding to the compactification radius in the Einstein frame.

\subsubsection*{Type IIA Solutions}
The (oxidized) Hopf T-dual solution of the six-dimensional type IIA theory is given by,
\begin{multline}
ds^2 = \frac{1}{(\lambda^2 + \mu^2)^{3/2}}[ - \cosh^2\rho d\tau^2 + d\rho^2  + \frac{\lambda^2}{\lambda^2+\mu^2}(dz'+\sinh\rho d\tau)^2]+ds^2(S^3_A)
\label{iia}
\end{multline}
where, $z = \lambda z'/(\lambda^2 + \mu^2)^{3/2} = Q_{NS} z'$. Once again, $z'$ is not necessarily identified  for \ref{iia} to be a solution of the type IIA theory. 
For convenience, we have defined $ds^2(S^3_A) \equiv (\lambda^2+\mu^2)^{-1/2}ds^2(S_B^3)$.

It is interesting to note that for real NS-NS and R-R charges we always find spacelike squashed warped dual solutions. This property is true even when we turn on non-zero constant values for the type IIB dilatons, as shown in appendix \ref{appb}.

Another interesting case given by a vanishing $Q_{RR}$ charge with $\mu = 0$. In this case the Hopf-T dualized solution is also a product of three-dimensional anti-de Sitter space and a sphere. This case will be of interest in the next section. On the other hand, if we turn off $Q_{NS}$, such that $\lambda = 0$ then the Hopf T-dual geometry becomes $AdS_2 \times S^1 \times S^3$.

The Ricci scalar of the $WAdS_3$ piece is given by
\be
R = -(\lambda^2 + \mu^2)^{1/2}(3\lambda^2 + 4 \mu^2)/2 = - (3Q_{NS}^2 + 4Q_{RR}^2)/(Q_{NS}^2 + Q_{RR}^2)^{7/4}
\ee
Notice that the region of parameter space where the original $AdS_3$ solution has small curvature is also that where the $WAdS_3$ has small curvature. On the other hand, the size of the compactified circle in the Hopf T-dual solution is given by $R_{IIA} = 1/R_{IIB}$ which is much smaller than the string scale when $R_{IIB} \gg 1$.


Finally, it is rather convenient to write the quantities $\ell$ and $\beta$ in terms of the new parameters $\lambda$ and $\mu$:
\bea
\ell^2 &=& \frac{12}{(\lambda^2 + \mu^2)^{1/2}(3\lambda^2 + 4\mu^2)} = \frac{6(Q_{NS}^2 + Q_{RR}^2)^{7/4}}{(3Q_{NS}^2 + 4Q_{RR}^2)} \label{l} \\
\beta^2 &=& \frac{\lambda^2}{\lambda^2 + \mu^2} = \frac{Q_{NS}^2}{Q^2_{NS}+Q^2_{RR}}
\eea

\section{Entropy of Six-Dimensional Black Strings}

In this section we compute the entropy of the black string solutions, with the fiber coordinate compactified, both from the six-dimensional gravitational point of view and from the point of view of the dual CFT.

\subsection{Six-dimensional Type IIB Black String}

We begin with the metric \ref{iib} which is the near horizon of the six-dimensional type IIB black string. This has been computed in a similar, but not identical, fashion in \cite{Strominger:1997eq,Kaloper:1998vw} (see also \cite{Horowitz:1996fn,Myung:1999pm}). The four dimensional area of the black string is given by the volume of the fixed $\rho$ and $\tau$ surface
\be
A_4 = \frac{16\pi^2}{(\lambda^2 + \mu^2)^2} L
\ee
where $L = \int dz$ is the coordinate length of the string, which is infinite in extent. In order to regularize this infinity we will identify the direction $z \sim z + 2\pi\alpha$ such that $L = 2\pi\alpha$. Strictly speaking, since we are identifying $z$, we are actually dealing with a five dimensional black hole. Thus, the Bekenstein-Hawking entropy of this solution is given by\footnote{We are using the following notation. Newton's constants with no tilde i.e. $G_d$ are those of the type IIB theory, whereas those with a tilde, i.e. $\tilde{G}_d$ are those of the type IIA theory.},
\be
S_{BH}^{IIB} = \frac{A_4}{4G_6}  = \frac{8\pi^3\alpha}{G_6(\lambda^2 + \mu^2)^2}
\ee
Now we can use our knowledge of $AdS_3/CFT_2$ to compute the entropy once again, from the point of view of the dual conformal field theory.  Notice that with $z$ identified, the $AdS_3$ metric is a self-dual metric with anti-de Sitter length
\be
\ell = \frac{2}{\sqrt{\lambda^2 + \mu^2}}
\ee
The central charge is given by Brown-Henneaux to be
\be
c^B_L = \frac{3\ell}{2G_3} = \frac{3}{G_3\sqrt{\lambda^2+\mu^2}}\label{cbhbi}
\ee
and the entropy of the CFT is given by:
\be
S_{CFT} = \frac{\pi^2 \ell c^B_L T_L}{3}
\ee
The left moving temperature $T_L$ is determined by  the periodicity of $z$ and is found to be $T_L = \frac{\alpha}{2\pi \ell}$. The six-dimensional Newton's constant is related to the three-dimensional Newton's constant by $G_6 = Vol(S_B^3)G_3$ so that
\be
G_3 = \frac{(\lambda^2+\mu^2)^{3/2} G_6}{16\pi^2} \label{g3}
\ee
Assembling all the pieces leaves us with an entropy for the dual CFT of the form:
\be
S_{CFT} = \frac{8\pi^3\alpha}{G_6(\lambda^2 + \mu^2)^2} = S_{BH}^{IIB}
\ee
and thus the entropies of the black-string with an $AdS_3\times S_B^3$ near horizon and the dual CFT match.

\subsubsection*{$c^B_L$ in the Ten-Dimensional Frame}

For later convenience we also write down the form of $c^B_L$ in terms of $G_6$ and the charges
\be
c^B_L = \frac{48\pi^2(Q_{RR}^2 + Q^2_{NS})}{G_6}\label{cbhb}
\ee

In the ten-dimensional theory we have that $16\pi G_{10} = (2\pi)^7\ell_s^8g_s^2$ where $G_{10} = (2\pi\ell_s)^4V_{T^4} G_6$ and $(2\pi\ell_s)^4V_{T^4}$ is the physical volume of the $T^4$. Furthermore, we have that $g_6 = g_s/\sqrt{V_{T^4}}$ which we should take into account even though we set $g_6 = 1$ by choosing $\phi_1 = \phi_2 = 0$. Thus, in the ten-dimensional frame
\be
c^B_L = 6 \times 2^4 (Q_{RR}^2 + Q^2_{NS})\frac{V_{T^4}}{g_s^2} = 6(\bar{Q}_{RR}^2 + \bar{Q}_{NS}^2) \label{tenb}
\ee
where we have defined $\bar{Q}_{RR} = 4Q_{RR}/g_6$ and $\bar{Q}_{NS} = 4Q_{NS}/g_6$. In the above form, the discrete character of the central charge is manifest. The central charge is quadratic in the charges bearing a similarity to the case $c_L = 6 Q_1 Q_5$ from the CFT found in the IR limit of the world volume gauge theory of the $D1$-$D5$ brane system \cite{Strominger:1996sh,Horowitz:1996fn}. In fact, if we switch off the $Q_{NS}$ we recover that precise situation.

\subsection{Six-Dimensional Hopf T-Dualized Type IIA Black String}

At this point we would like to apply our central charges \ref{cw} to account for the entropy of the Hopf T-dualized black string whose near horizon is given by $WAdS_3 \times S^3$ with metric \ref{iia}. The Bekenstein-Hawking entropy is found to be
\be
S_{BH}^{IIA} = \frac{A_4}{4 \tilde{G}_6} = \frac{4\pi^2 \lambda}{\tilde{G}_6(\lambda^2 + \mu^2)^{7/2}}L
\ee
where $L = \int dz'$. Once again, $L$ is infinite in extent and we will identify the fiber coordinate as $z' \sim z' + 2\pi\alpha'$ such that $L=2\pi\alpha'$. Then we have that our solution is a warped self-dual solution and we can compute the entropy of the dual CFT. The central charge is now given by the proposed value in \ref{cw}. In terms of $\lambda$ and $\mu$ it becomes
\be
c^A_L = \frac{3\lambda}{\tilde{G}_3(\lambda^2 + \mu^2)^{5/4}}\label{cwai}
\ee
The left moving temperate is easily found to be $T_L = \alpha'/2\pi\ell$ as before, where now $\ell$ is given in \ref{l}, and $\tilde{G}_3$ is given by
\be
\tilde{G}_3 = \frac{(\lambda^2 + \mu^2)^{9/4}\tilde{G}_6}{16\pi^2}\label{g3a}
\ee
Notice that \ref{g3} and \ref{g3a} are different since the volume of the three sphere is different in \ref{iib} and \ref{iia}.

Thus, the entropy of the CFT becomes
\be
S_{CFT} = \frac{\pi^2 \ell c^A_L T_L}{3} = \frac{8\pi^3\lambda\alpha'}{\tilde{G}_6(\lambda^2 + \mu^2)^{7/2}} = S_{BH}^{IIA}
\ee
Once again, the Bekenstein-Hawking entropy and the entropy coming from the dual CFT match.

\subsubsection*{$c^A_L$ in the Ten-Dimensional Frame}

In terms of $\tilde{G}_6$ and the charges $c^A_L$ is given by
\be
c^A_L = \frac{48\pi^2(Q_{NS}^2 + Q_{RR}^2) Q_{NS}}{\tilde{G}_6}\label{cwa}
\ee
We can express this in the ten-dimensional frame where we find
\be
c^A_L = 6(\bar{Q}_{NS}^2 + \bar{Q}_{RR}^2)Q_{NS}\label{tena}
\ee
One immediately notices that the central charge is now cubic in the charges, indicating a deviation from the type IIB situation. It would be interesting to further understand this property and we leave this for future work.

\section{Duality Chains}

From Brown-Henneaux we learned that quantum gravity in asymptotically $AdS_3$ spacetimes is holographically dual to a two-dimensional CFT, say $CFT_{B}$. Furthermore, the above results suggest that asymptotically $WAdS_3$ is holographically dual to another two-dimensional CFT, say $CFT_{A}$. One may therefore ask what the relation between these two CFTs is, in light of the Hopf T-dualization.

\subsection{Hopf T-Dualization along the $AdS_3$ fiber}

In this case we can extend the arrow diagram of \ref{arrow} to
\be
CFT_B \longleftrightarrow AdS_3\times S_B^3 \longleftrightarrow WAdS_3 \times S_A^3 \longleftrightarrow CFT_A
\ee
The precise Hopf T-duality relates warped self-dual solutions with unwarped self-dual solutions. We argued that self-dual solutions are in fact states in a two-dimensional CFT with the right moving temperature turned off. 


The ansatz \ref{dimred} we have used for the dimensional reduction of the six dimensional metric is such that $\int dz d^5 x \sqrt{-\hat{g}}\hat{R} = 2\pi  \int d^5x \sqrt{g}(R+...)$, where the hat denotes six-dimensional quantities \cite{pope,Duff:1998us}. This ensures that the six-dimensional Newton's constant remains unchanged before and after the Hopf T-dualization, i.e. $G_6 = \tilde{G}_6$, so long as the $z$ coordinate retains its periodicity and $G_5$ retains its value. It follows from \ref{cbhb} and \ref{cwa} that
\be
\frac{c^A_L}{c^B_L} = Q_{NS} \frac{G_6}{\tilde{G}_6} = Q_{NS}
\ee
as expected from \ref{tenb} and \ref{tena}.

Now, recalling that the duality chain relates two self-dual solutions, one warped and the other unwarped, we can compare the entropy of each and see that they match. This happens because the ratio of temperatures is given by the change in periodicity of the fiber coordinate given in \ref{iia}. So we have
\be
S_{CFT}^B = \frac{\pi^2 \ell_B T_L^B c_L^B}{3} = \frac{\pi^2 \ell_A T_L^A c_L^A}{3} = S_{CFT}^A
\ee

Thus thermal states with vanishing right-moving temperature in $CFT_B$ are mapped to thermal states with vanishing right-moving temperature in $CFT_A$ with the same entropy. The fact that they have the same entropy may not come as a surprise given that it is emphasized in \cite{Duff:1998cr} that the area of the horizon is preserved under Hopf T-dualization and thus the Bekenstein-Hawking entropy itself is preserved\footnote{It is also interesting to note the discussion of \cite{Horowitz:1993wt} on the invariance of Hawking entropy under certain string dualities.}.

One can also consider the above duality chain with vanishing $Q_{RR}$ charge, such that $\mu = 0$. In this case, the map is between $AdS_3 \times S^3_B$ and $AdS_3 \times S^3_A$ and thus the central charges are known. Once again, we find that the ratio of the two Brown-Henneaux central charges $c^A_L$ and $c_L^B$ is $Q_{NS}$; thus the map is between two different CFTs.

\subsubsection*{Multiplicity of States}

It is somewhat useful to consider the partition function of a 2d CFT with central charge $c^B_L$ for the left-movers
\be
Z_L = \sum^{\infty}_{N_L=1} d(N_L) e^{-\beta_L E_L}\label{part}
\ee
where $d(N_L)$ is the multiplicity of left-moving states and $N_L$ is the $\bar{L}_0$ eigenvalue. Given a two-dimensional unitary CFT, Cardy's formula tells us that for $N_L \gg 1$ the multiplicty of states goes as $d(N_L) \sim e^{2\pi \sqrt{c_LN_L/6}}$ \cite{Cardy:1986ie}.

Recall that under the Hopf T-duality the effective temperatures of the type IIB and type IIA solutions with $z$ and $z'$ identified are in fact related by $ \ell_B T^B_L = Q_{NS} \ell_A T^A_L$ (since they are proportional to the periodicities of $z$ and $z'$). Using further that $c_L^A/c_L^B = Q_{NS}$ and $N_L \sim c_L (T_L \ell)^2$ we obtain $N_L^A/N^B_L = 1/Q_{NS}$.
Performing the transformation
\be
c^B_L \rightarrow c^A_L = Q_{NS} c^B_L , \quad N^B_L \rightarrow  N^A_L = N^B_L/Q_{NS}
\ee
on $d(N_L)$, we find that the multiplicity of states for large $N^A_L$ with $c^A_L$ is equal to that for large $N^B_L$ with $c^B_L$. Thus, we see that it is at least conceivable that the two partition functions of $CFT_B$ and $CFT_A$ can, and in fact should, be equal under the Hopf T-dualization even if $c_L^A \neq c_L^B$\footnote{From the stringy point of view, it is possible that the effective CFTs coming from the long string sectors match \cite{Maldacena:1996ds}.}.

\subsection{Hopf T-Dualization along the $S^3_B$ fiber}

As demonstrated in \cite{Duff:1998cr} one can also perform the Hopf T-duality along the fiber coordinate\footnote{Recall the metric of a three-sphere can be written as a fibration over $S^2$ where the fiber coordinate $x$ is an $S^1$ with $x \sim x + 4\pi$.} of the $S^3_B$ which results in the metric \cite{Duff:1998cr}
\begin{multline}
ds^2 = \frac{1}{(\lambda^2+\mu^2)^{3/2}}\left[ -\cosh^2\rho d\tau^2 + d\rho^2 + (dz + \sinh\rho d\tau)^2\right]  \\ + \frac{1}{(\lambda^2+\mu^2)^{3/2}}\left[ d\theta^2 + \sin\theta^2d\phi^2 + \frac{\lambda^2}{(\lambda^2 + \mu^2)}(dx' + \cos\theta d\phi)^2 \right]
\end{multline}
where $x = Q_{NS} x'$ and $x' \sim x' + 4\pi/Q_{NS}$. This is the product of $AdS_3$ with a squashed sphere $s^3_A$ and thus is dual to a two-dimensional CFT. The anti-de Sitter length can be easily read off and is given by $\ell = 2/(\lambda^2 + \mu^2)^{3/4}$. Thus the Brown-Henneaux central charge is given by
\be
\bar{c}^A_L = \frac{3}{\tilde{G}_3(\lambda^2+\mu^2)^{3/4}} = \frac{3}{\tilde{G}_6}\frac{16\pi^2}{(\lambda^2+\mu^2)^{2}} = \frac{48\pi^2(Q_{RR}^2 + Q^2_{NS})}{\tilde{G}_6}
\ee
where we have used that $\tilde{G}_6 = Vol(s^3_A)\tilde{G}_3$.

Note that $\bar{c}^A_L$ is precisely the same central charge as the original one in \ref{cbhb}, when expressed in terms of $\tilde{G}_6$. Thus, Hopf T-Dualizing along the $S^3_B$ fiber maps states with vanishing right-moving temperature to states with vanishing right-moving temperature with the same entropy. Furthermore, the CFTs have the same central charge.


\subsection{Double Hopf T-Dualization}

The final case of interest is given by first Hopf T-dualizing along the $AdS_3$ fiber coordinate and then Hopf T-dualizing along the $S^3_A$ fiber coordinate. The final metric is given by \cite{Duff:1998cr}
\begin{multline}
ds^2 = \frac{1}{(\lambda^2 + \mu^2)^2}\left[ -\cosh^2\rho d\tau^2 + d\rho^2 + \frac{\lambda^2}{(\lambda^2 + \mu^2)}(dz'+\sinh\rho d\tau)^2 \right] \\
+ \frac{1}{(\lambda^2 + \mu^2)^2}\left[ d\theta^2 + \sin\theta^2d\phi^2 + \frac{\lambda^2}{(\lambda^2 + \mu^2)}(dx' + \cos\theta d\phi)^2  \right]
\end{multline}
this case, we have a product of $WAdS_3$ and a squashed sphere $s^3_B$ which we have proposed to be dual to a two-dimensional CFT. The central charge can be computed using \ref{cw} and it becomes
\be
\bar{c}^B_L = \frac{3\lambda}{G_3(\lambda^2 + \mu^2)^{3/2}} = \frac{3}{G_6}\frac{16\pi^2\lambda}{(\lambda^2+\mu^2)^{7/2}} = \frac{48\pi^2 Q_{NS}(Q_{NS}^2 + Q_{RR}^2)}{G_6}
\ee
where we have used that $G_6 = Vol(s^3_B)G_3$.

Once again, note that $\bar{c}^B_L$ is precisely the same central charge as the original one in \ref{cwa}, when expressed in terms of $G_6$. Thus, Hopf T-Dualizing type IIA $WAdS_3\times S^3_A$ solution along the $S^3_A$ fiber maps states with vanishing right-moving temperature to states with vanishing right-moving temperature with the same entropy and the CFTs have the same central charge.


\section{Discussion}

Following previous work \cite{Anninos:2008fx,Compere:2008cw,Compere:2007in}, we have proposed that three-dimensional spacelike warped anti-de Sitter space, as a solution to Einstein theories of gravity, is dual to a two dimensional CFT with central charges
\be
c_L = c_R = \frac{\sqrt{3}(4-\beta^2)^{1/2} \beta \ell}{2G_3} 
\ee
for all values of $\beta^2$ in the range $\beta^2 \in [0,4]$.

Our evidence is based mostly on consistency checks applied to the Bekenstein-Hawking entropy of stretched black holes and warped self-dual solutions. We have least evidence for the value of $c_R$ and in particular for $c_R$ when $\beta^2<1$; the only non-pathological squashed solutions we have are warped self-dual solutions which are sensitive only to $c_L$. If the diffeomorphism anomaly vanishes in the boundary theory, knowing $c_L$ is enough to learn the value of $c_R$. Also, that the asymptotic symmetry group of the squashed black hole gives $c_R$ up to a sign seems encouraging \cite{Compere:2007in}.

We have furthermore used the left moving central charge to account for the Bekenstein-Hawking entropy of the Hopf T-dual black strings with a $WAdS_3\times S^3$ near horizon discovered in \cite{Duff:1998cr}. It would thus be interesting to see if these central charges can be computed by studying the asymptotic symmetries of such theories.

We have completed the duality chain discovered in \cite{Duff:1998cr} as follows
\be
CFT_B \longleftrightarrow AdS_3\times S^3 \longleftrightarrow WAdS_3 \times S^3 \longleftrightarrow CFT_A
\ee
The above map is between thermal states of the two CFTs with vanishing right-moving temperature and equal entropy. It was also found that Hopf T-dualizing along the $S^3$ fiber coordinate of either $WAdS^3\times S^3$ or $AdS^3 \times S^3$ does not affect the central charge of the dual theory. In such cases the map is between thermal states with vanishing right-moving temperature and equal entropy in CFTs with equal central charges.

It may seem more intuitive that $CFT_A$ and $CFT_B$ are in fact identical; however, we find no direct evidence for this. On the contrary, when $Q_{RR} = 0$ and the Hopf T-duality is a map between two $AdS_3$s, the Brown-Henneaux central charges are still different. It is, however, already an interesting feature of the Hopf T-duality that the degrees of freedom of the dyonic black string with a $WAdS_3\times S^3$ near horizon conspire to those of a known two-dimensional CFT, namely the one dual to the $AdS_3\times S^3$ type IIB solution. One possibility is that one must consider the long string sector, which is described by an effective CFT with rescaled central charge and effective temperature, in order to understand how the theories match microscopically.

It is well known that the dyonic black string can been obtained from an intersecting NS-NS 1-brane and NS-NS 5-brane in the ten dimensional picture \cite{Strominger:1996sh}. Other D-brane configurations can also be obtained from various string dualities. In the same spirit, one would also like to have a D-brane interpretation of the Hopf T-dual dyonic string with $WAdS_3 \times S^3$ near horizon in order to understand the world volume theory. One challenge is to better understand how to construct CFTs with when we have both R-R and NS-NS charges turned on.




There is another limit of interest. In particular, if we have vanishing $Q_{NS}$ charge, the six-dimensional Hopf T-dual solution becomes $AdS_2\times S^1 \times S^3$. In this case the duality chain becomes
\be
CFT_B \longleftrightarrow AdS_3\times S^3 \longleftrightarrow AdS_2 \times S^1 \times S^3
\ee
It would be interesting to see whether we can learn something about $AdS_2$ and its dual CFT using the above relation \cite{Strominger:1998yg,Hartman:2008dq}. For instance, the above chain may imply that $AdS_2\times S^1$ resembles a chiral half of a two-dimensional CFT with vanishing $T_R$ rather than conformal quantum mechanics.

In all known cases where $WAdS_3$ (or identifications thereof) appears in Einstein theories, it is only found to be both stretched and with matter content free of pathologies for the near horizon of the extremal Kerr black hole at fixed polar angle \cite{Bengtsson:2005zj,Bardeen:1999px}. It would be of interest to understand the precise relation of the proposed central charges and the dual theory of extremal Kerr black holes that has been recently discussed \cite{ghss,sstrings}. It seems that the proposed $c_L$ should be the relevant central charge at each fixed polar angle, and once integrated over appropriately one should retrieve the extremal Kerr black hole entropy \cite{tom}. For instance, given the extremal Kerr metric \ref{kerr}, we can express the left moving central charge as,
\be
c_L = \frac{6\sqrt{2}}{G_3}\frac{M\sin\theta}{\sqrt{(3+\cos{2\theta})}}
\ee
The three dimensional Newton's constant can be related to the four dimensional one for a polar slice between $\theta$ and $\theta + d\theta$ as
\be
\frac{1}{G_3} = \frac{1}{G_4} d\theta \sqrt{g_{\theta\theta}} = \frac{1}{G_4} \sqrt{2} M d\theta \sqrt{\frac{1+\cos^2\theta}{2}}
\ee
Curiously, and perhaps interestingly, integrating over $\theta$ gives
\be
c_L^{total} = 12 J
\ee
This is the central charge obtained for the extremal Kerr black hole in \cite{ghss}.

Finally, given that regular matter supports squashed $WAdS_3$ in all other known cases, one would like to construct a model with non-pathological matter content supporting stretched $WAdS_3$.

\section*{Acknowledgements}
This work was partially funded by an DOE grant DE-FG02-91ER40654. It has been a great pleasure to discuss this work with F. Denef, M. Esole, T. Hartman and A. Strominger.

\appendix

\section{$AdS_3$ Isometries}
\label{appa}

The isometries of this solution are given by $SL(2,\mathbb{R})_L \times SL(2,\mathbb{R})_R$. The $SL(2,\mathbb{R})_L$ isometries are given by
\begin{eqnarray}
J_1 &=& {\frac{2 \sinh{z}}{\cosh{\rho}} \partial_{\tau}-2\cosh{z} \partial_{\rho}+2\tanh{\rho} \sinh{z} \partial_{z}} \\
J_2 &=& 2\partial_{z} \\
J_0 &=& {-\frac{2 \cosh{z}}{\cosh{\rho}} \partial_{\tau}+2\sinh{z} \partial_{\rho}-2\tanh{\rho} \cosh{z} \partial_{z}}
\end{eqnarray}
These satisfy the algebra $[J_1,J_2]=2J_0$, $[J_0,J_1]=-2J_2$ and $[J_0,J_2]=2J_1$. The $SL(2,\mathbb{R})_R$ isometries are given by
\begin{eqnarray}
\tilde{J}_1 &=&  {2\sin{\tau} \tanh{\rho} \partial_{\tau} - 2\cos{\tau} \partial_{\rho}+\frac{2\sin{\tau}}{\cosh{\rho}} \partial_{z} }  \\
\tilde{J}_2 &=&  {-2\cos{\tau} \tanh{\rho} \partial_{\tau}-2\sin{\tau} \partial_{\rho} - \frac{2\cos{\tau}}{\cosh{\rho}} \partial_{z}  }  \\
\tilde{J}_0 &=&  2\partial_{\tau}
\end{eqnarray}
These satisfy the algebra $[\tilde{J}_1,\tilde{J}_2]=2\tilde{J}_0$, $[\tilde{J}_0,\tilde{J}_1]=-2\tilde{J}_2$ and $[\tilde{J}_0,\tilde{J}_2]=2\tilde{J}_1$.

\section{Hopf T-dual Solutions with non-zero $\phi_1$ and $\phi_2$}
\label{appb}
In order to answer whether there are stretched spacelike warped solutions we must take a close look at the equations of motion. The six-dimensional truncated type IIB Lagrangian is given by:
\begin{multline}
e^{-1}\mathcal{L}_{6B} = R - \frac{1}{2}(\partial\phi_1)^2 - \frac{1}{2}(\partial\phi_2)^2 -\frac{1}{2}e^{2\phi_1}(\partial \chi_1)^2 -\frac{1}{2}e^{2\phi_2}(\partial \chi_2)^2 \\ - \frac{1}{12}e^{-\phi_1-\phi_2}(F^{NS}_{(3)})^2 - \frac{1}{12}e^{\phi_1-\phi_2}(F^{RR}_{(3)})^2+\chi_2 dA^{NS}_{(2)}\wedge dA^{RR}_{(2)}
\end{multline}
The equations of motion are solved by the following three-forms:
\bea
F^{NS}_{(3)} &=& \lambda \varepsilon(AdS_3) + \lambda \varepsilon(S^3)\\
F^{RR}_{(3)} &=& \mu \varepsilon(AdS_3) + \mu \varepsilon(S^3)
\eea
where $\varepsilon(X)$ denotes the volume form of $X$. The metric is given by:
\be
ds^2_6 = ds^2(AdS_3) + ds^2(S^3)
\ee
with Ricci tensors,
\be
R_{\mu \nu} = -\frac{1}{2}(e^{-\phi_1-\phi_2}\lambda^2+e^{\phi_1-\phi_2}\mu^2)g_{\mu \nu}, \quad R_{m n} = \frac{1}{2}(e^{-\phi_1-\phi_2}\lambda^2+e^{\phi_1-\phi_2}\mu^2)g_{mn}
\ee
The equations of motion for the scalar fields are trivially satisfied once we choose the three-forms to be self-dual, i.e. the coefficients of the $AdS_3$ and $S^3$ pieces to be equal. The Greek indices are those of $AdS_3$ and the Latin indices are those of the $S^3$.

For completeness, we also give the six-dimensional type IIA Lagrangian below
\begin{multline}
e^{-1} \mathcal{L}_{6A} =  R - \frac{1}{2}(\partial \phi_1)^2 - \frac{1}{2}(\partial \phi_2)^2  \\ - \frac{1}{48}e^{\phi_1/2 - 3\phi_2/2}\left(F_{(4)}\right)^2 - \frac{1}{12}e^{-\phi_1 - \phi_2}\left(F_{(3)}\right)^2 - \frac{1}{4}e^{3\phi_1/2 - \phi_2/2}\left(F_{(2)}\right)^2
\end{multline}

\subsection{Hopf T-Dual Solution}

Having obtained the more general IIB solution we can Hopf T-dualize along the $AdS_3$ fiber coordinate to obtain a IIA solution. The explicit IIB metric in the fibrated coordinates is given by:
\be
ds^2 = \frac{1}{(e^{-\phi_1-\phi_2}\lambda^2+e^{\phi_1-\phi_2}\mu^2)}[ - \cosh^2\rho d\tau^2 + d\rho^2+ (dz+\sinh\rho d\tau)^2] + ds^2\left(S^3\right)
\ee
Reducing along the fiber coordinate $z$ with the following ansatz for the decomposition of the metric:
\be
ds^2_6 = e^{-\varphi/\sqrt{6}}ds^2_5 + e^{\sqrt{3/2}\varphi}(dz + \mathcal{A}_{(1)})^2\label{dimred}
\ee
we obtain
\bea
e^{\varphi/\sqrt{6}} &=& (e^{-\phi_1-\phi_2}\lambda^2+e^{\phi_1-\phi_2}\mu^2)^{-1/3} \equiv \Delta^{-1/3}\\
\mathcal{F}_{(2)} &\equiv& d\mathcal{A}_{(1)} =  \cosh\rho d\rho d\tau = \Sigma_{(2)}
\eea
and the five dimensional metric
\be
ds^2_5 = \Delta^{-4/3}\left(-d\tau^2 + \cosh^2\rho d\tau^2 \right)  + \Delta^{-1/3} ds^2(S^3)
\ee
Once in the five-dimensional space we can Hopf T-dualize the IIB solution using the rules obtained in \cite{Duff:1998cr}. The five-dimensional field content of the the IIB theory is given by
\bea
F^{NS}_{(3)} &=& \lambda \varepsilon(S^3), \quad F^{NS}_{(2)1} = \frac{\lambda}{{\Delta}^{3/2}}\Sigma_{(2)}\\
F^{RR}_{(3)} &=& \mu \varepsilon(S^3), \quad F^{RR}_{(2)1} = \frac{\mu}{{\Delta}^{3/2}}\Sigma_{(2)}\\
\mathcal{F}_{(2)} &=& \Sigma_{(2)}
\eea
The T-dual field content in the five-dimensional IIA theory become
\bea
F_{(3)} &=& \lambda \varepsilon(S^3), \quad \mathcal{F}'_{(2)} = \frac{\lambda}{{\Delta}^{3/2}}\Sigma_{(2)}\\
F_{(3)1} &=& -\mu \varepsilon(S^3), \quad F_{(2)} = \frac{\mu}{{\Delta}^{3/2}}\Sigma_{(2)}\\
F_{(2)1} &=& \Sigma_{(2)}
\eea
The scalars in the type IIA theory become
\bea
\phi_1' &=& \frac{1}{4}\left( -\sqrt{6}\varphi + 3\phi_1 - \phi_2 \right) \\
\phi_2' &=& \frac{1}{4}\left( -\sqrt{6}\varphi - \phi_1 + 3\phi_2 \right) \\
\varphi' &=&  \frac{1}{4}\left( -2\varphi - \sqrt{6}(\phi_1 + \phi_2) \right)
\eea
Finally we can oxidize back to the six-dimensional type IIA solution
\be
ds^2_6 = e^{-\varphi'/\sqrt{6}}ds^2_5 + e^{\sqrt{3/2}\varphi'}(dz + \mathcal{A}'_{(1)})^2
\ee
If we express this more explicitly,
\begin{multline}
ds^2_6 = e^{-\varphi'/\sqrt{6}}\Delta^{-4/3}\left[ \left( - \cosh^2\rho d\tau^2 + d\rho^2  \right)  +
\lambda^2 e^{4\varphi'/\sqrt{6}}\Delta^{-5/3} (dz' + \mathcal{A}_{(1)})^2 \right] \\ + e^{-\varphi'/\sqrt{6}}\Delta^{-1/3}ds^2(S^3)
\end{multline}
where $z = \lambda\Delta^{-3/2}z'$. One can easily check that we recover the metric \ref{iia} once we set $\phi_1 = \phi_2 = 0$. The warping factor can be expressed more explicitly:
\be
\lambda^2 e^{4\varphi'/\sqrt{6}}\Delta^{-5/3} = \frac{\lambda^2 e^{-(\phi_1+\phi_2)}}{(e^{-(\phi_1+\phi_2)}\lambda^2+e^{\phi_1-\phi_2}\mu^2)}
\ee
In conclusion, we still obtain a squashed metric for all constant values of the scalars $\phi_1$ and $\phi_2$.

\end{document}